\documentclass[10pt,conference]{IEEEtran}
\usepackage[cmex10]{amsmath}
\usepackage{amssymb}
\usepackage{amsfonts}
\usepackage{wrapfig}
\usepackage{graphicx}
\usepackage{enumerate}
\usepackage{xspace}

\newcommand{\Icf}{\ti{IC4-PD}\xspace}
\newcommand{\ict}{\ti{IC3}\xspace}
\newcommand{\icf}{\ti{IC4}\xspace}
\newcommand{\icfl}{\ti{IC4}$^*$\xspace}
\newcommand{\qp}{\ti{QUIP}\xspace}
\newcommand{\Cr}[1]{$\mi{ConvRate}(#1)$\xspace}

\newcommand{\pnt}[1]{{\mbox{\boldmath $#1$}}}

\newcommand{\nGz}[2]{$G_{non-\{z\}}$}

\newcommand{\mi}[1]{\mathit{#1}}
\newcommand{\ti}[1]{\textit{#1}}
\newcommand{\tb}[1]{\textbf{#1}}

\newcommand{\ttt}{\>\>\>}
\newcommand{\tttt}{\>\>\>\>}

\newcommand{\Tt}{\>\>}

\newcommand{\Comment}[1]{}

\newcommand{\ks}{\mbox{$\xi$}\xspace}

\newcommand{\Abs}[2]{\mbox{$\mathbb{#1}_{#2}$}}

\mathchardef\mhyphen="2D

\newcommand{\s}[1]{\mbox{$\{#1\}$}}
\newcommand{\di}[1]{\mbox{$\mi{Diam}(#1)$}}
\newcommand{\gs}[1]{\mbox{$|#1$-$\mi{states}|$}\xspace}

\begin{document}

\title{Improving Convergence Rate Of IC3}

\author{\IEEEauthorblockN{Eugene Goldberg} 
\IEEEauthorblockA{
eu.goldberg@gmail.com}}

\maketitle

\begin{abstract}
\ict, a well-known model checker, proves a property of a transition
system \ks by building a sequence of formulas $F_0,\dots,F_k$. Formula
$F_i$, $0 \leq i \leq k$ over-approximates the set of states reachable
in at most $i$ transitions.  The basic algorithm of \ict cannot
guarantee that the value of $k$ never exceeds the reachability
diameter of \ks.  We describe an algorithm called \icf that gives such
a guarantee. (\icf stands for ''\ict + Improved Convergence'').  One
can argue that the \ti{average} convergence rate of \icf is better
than for \ict as well.  Improving convergence can facilitate some
other variations of the basic algorithm.  As an example, we describe a
version of \icf employing \ti{property decomposition}. The latter
means replacing an original (strong) property with a conjunction of
weaker properties to prove by \icf. We argue that addressing the
convergence problem is important for making the property decomposition
approach work.
\end{abstract}

\section{Introduction}
\ict is a model checker~\cite{ic3} that has become very popular due to
its high scalability. Let \ks be a transition system and $P$ be a
safety property of \ks. \ict builds a sequence of formulas
$F_0,\dots,F_k$ where $F_i$ over-approximates the set of states
reachable from an initial state of \ks in at most $i$ transitions.
Property $P$ is proved when $F_i$ becomes an inductive invariant of
\ks for some $0 \leq i \leq k$.

One of the reasons for high performance of \ict is that the value of
$k$ above is typically much smaller than \di{\ks} (i.e. the
reachability diameter of \ks). So, on average, \ict converges to an
inductive invariant much faster than an RA-tool (where RA stands for
``reachability analysis''). Interestingly, the \ti{worst case}
behavior of an RA-tool and \ict is quite different from their average
behavior. Namely, \ict cannot guarantee that $k$ never exceeds
\di{\ks}.  We introduce a modification of \ict called \icf that fixes
the problem above. (\icf stands for ``\ict + Improved
Convergence''). On one hand, \icf has the same worst case behavior as
an RA-tool. On the other hand, the \ti{average} convergence rate of
\icf is arguably better than that of \ict as well.

The main difference between \icf and \ict is as follows. \ict checks
if formula $F_k$ is an inductive invariant by ``pushing'' the clauses
of $F_k$ to $F_{k+1}$. If every clause of $F_k$ can be pushed to
$F_{k+1}$, the former is an inductive invariant. Otherwise, there is
at least one clause $C \in F_k$ that cannot be pushed to $F_{k+1}$.
In this case, \ict moves on re-trying to push $C$ to $F_{k+1}$ when
new clauses are added to $F_k$. In contrast to \ict, \icf applies
extra effort to push $C$ to $F_{k+1}$. Namely, it derives new
inductive clauses to exclude states that prevent $C$ from being pushed
to $F_{k+1}$. This extra effort results either in successfully pushing
$C$ to $F_{k+1}$ or in \ti{proving} that $C$ is ``\ti{unpushable}''.

The proof of unpushability consists of finding a \ti{reachable} state
\pnt{s} that satisfies formula $F_{k+1}$ and falsifies clause $C$. The
existence of \pnt{s} means that $F_k$ \ti{cannot} be turned into an
inductive invariant by adding more clauses. Thus, semantically, the
difference between \icf and \ict is that the former starts building a
new over-approximation $F_{k+1}$ \ti{only} after it proved that adding
one more time frame is \ti{mandatory}. Operationally, \icf and \ict
are different in that \icf generates a small set of reachable states.

An appealing feature of \ict is its ability to generate
property-specific proofs. So it seems natural to decompose a hard
property $P$ into a conjunction $P_1 \wedge \dots P_m$ of weaker
properties and then generate $m$ property-specific proofs for $P_i$.
However, the convergence issues of \ict are arguably more pronounced
for weak properties (see Subsection~\ref{ssec:weak_prop}).  So, to
make property decomposition work, one should use \icf rather than \ict
to prove properties $P_i$. In this paper, we describe a variation of
\icf employing property decomposition.

At the time of writing the first version of the paper we were not
aware of \qp, a version of \ict published at \cite{quip}. We fix this
omission and describe the relation between \icf and \qp in
Subsection~\ref{ssec:quip}. \qp more aggressively than the basic \ict
pushes clauses to future time frames and generates reachable states as
a proof that a clause cannot be pushed. However, no relation of \qp's
good performance with improvement of its convergence rate has been
established either theoretically or experimentally.

The contribution of this paper is as follows. First, we show the
reason why \ict has a poor upper bound on the convergence rate
(Section~\ref{sec:conv_prob}). Second, we formulate a new version of
\ict called \icf (Section~\ref{sec:fix_prob}) that is meant for fixing
this problem. In particular, we show that \icf indeed has a better
upper bound than \ict (Section~\ref{sec:bet_conv_rate}).  We also give
an estimate of the number of reachable states \icf has to generate
(Section~\ref{sec:count_states}).  Third, we discuss arguments in
favor of \icf (Section~\ref{sec:args_in_favor}). Fourth, we describe
\Icf, a version of \icf meant for solving hard problems by property
decomposition (Section~\ref{sec:ic4_pd}).

\section{A Brief Overview Of \ict}
\label{sec:overview}
 Let $I$ and $T$ be formulas\footnote{We assume that all formulas are
   propositional and are represented in CNF (conjunctive normal form)}
 specifying the initial states and transition relation of a transition
 system \ks respectively.  Let $P$ be a formula specifying a safety
 property of \ks. \ict proves $P$ by building a set of formulas
 $F_0,\dots,F_k$.  Here formula $F_i$, $0 \leq i \leq k$ depends on
 the set of state variables of $i$-th time frame (denoted as $S_i$)
 and over-approximates the set of states\footnote{A state is an
   assignment to the set of state variables.} reachable in at most $i$
 transitions.  That is every state reachable in at most $i$
 transitions is an $F_i$-state
\footnote{Given a formula $H(S)$, a state \pnt{s} is said to be an
  $H$-state if $H(\pnt{s})=1$.}.

\ict builds formula $F_k$ as follows. Formula $F_0$ is always equal to
$I$.  Every formula $F_k$, $k > 0$ is originally set to $P$. (So $F_k
\rightarrow P$ is always true because the only modification applied to
$F_k$ is adding clauses.) Then \ict tries to exclude every $F_k$-state
that is a predecessor of a bad state\footnote{Given a property $P$, a
  $\overline{P}$-state is called a bad state.} i.e. a state \pnt{s}
that breaks $F_k \wedge T \rightarrow P'$.  Here $T$ is a short for
$T(S_k,S_{k+1})$ and $P'$, as usual, means that $P$ depends on
\ti{next-state variables} i.e. those of $S_{k+1}$. Exclusion of
\pnt{s} is done by derivation of a so-called inductive clause $C$
falsified by \pnt{s}. Adding $C$ to $F_k$ excludes \pnt{s} from
consideration. (If \pnt{s} cannot be excluded, \ict generates a
counterexample.)

One of the properties of formulas $F_i$ maintained by \ict is $F_i
\rightarrow F_{i+1}$.  To guarantee this, \ict maintains two stronger
properties of $F_i$: a) $\mi{Clauses}(F_{i+1}) \subseteq
\mi{Clauses}(F_i)$ and b) $F_i \neq F_{i+1}$ implies that $F_i
\not\equiv F_{i+1}$.  That is the set of clauses of $F_i$ contains all
the clauses of $F_{i+1}$ and the fact that $F_i$ contains at least one
clause that is not in $F_{i+1}$ means that $F_i$ and $F_{i+1}$ are
logically inequivalent. Since every formula $F_i$ implies $P$, one
cannot have more than \gs{P} different formulas $F_0, \dots, F_k$.
That is if the value of $k$ exceeds \gs{P}, there should be two
formulas $F_{i-1}$, $F_i$, $i < k$ such that $F_{i-1} = F_i$.  This
means that $F_{i-1}$ is an inductive invariant and property $P$ holds.

\section{Convergence Rate Of \ict And Clause Pushing}
\label{sec:conv_prob}
We will refer to the number of time frames one has to unroll before
proving property $P$ as the \tb{convergence rate}. We will refer to
the latter as \Cr{P}. As we mentioned in Section~\ref{sec:overview},
an upper bound on \Cr{P} of the basic version of \ict formulated
in~\cite{ic3} is \gs{P}. Importantly, the value of \gs{P} can be much
larger than \di{\ks} (i.e. the reachability diameter of \ks).  Of
course, on average, \Cr{P} of \ict is much smaller than \di{\ks}, let
alone \gs{P}. However, as we argue below, a poor upper bound on \Cr{P}
is actually \ti{a symptom of a problem}.

Recall that formula $F_k$ specifies an over-approximation of the set of
states reachable in at most $k$ transitions. So, it cannot exclude a
state \pnt{s} reachable in $j$ transitions where $j \leq k$.  (That is
such a state \pnt{s} cannot falsify $F_k$.) On the other hand, $F_k$
may exclude states reachable in \ti{at least} $k+1$ transitions or
more.

Suppose \ict just finished constructing formula $F_k$.  At this point
$F_k \wedge T \rightarrow P'$ holds i.e. no bad state can be reached
from an $F_k$-state in one transition. After constructing $F_k$, \ict
invokes a procedure for pushing clauses from $F_k$ to $F_{k+1}$. In
particular, this procedure checks for every clause $C$ of $F_k$ if
implication $F_k \wedge T \rightarrow C'$ holds.  We will refer to
this implication as the \tb{pushing condition}. If the pushing
condition holds for clause $C$, it can be pushed from $F_k$ to
$F_{k+1}$.  If the pushing condition holds for every
clause\footnote{In reality, since both $F_k$ and $F_{k+1}$ contain the
  clauses of $P$, only the inductive clauses of $F_k$ added to
  strengthen $P$ are checked for the pushing condition.} of $F_k$,
then $F_k \wedge T \rightarrow F'_k$ and $F_k$ is an inductive
invariant.

Suppose that the pushing condition does not hold for a clause $C$ of
$F_k$.  Below, we describe two different reasons for the pushing
condition to be broken. \ict does not try to identify which of the
reasons takes place. This feature of \ict is the cause of its poor
upper bound on \Cr{P}. Moreover, intuitively, this feature should
affect the \ti{average} value of \Cr{P} as well.

The first reason for breaking the pushing condition is that clause $C$
excludes a state \pnt{s} that is \ti{reachable} in $(k+1)$-th time
frame from an initial state.  In this case, formula $F_k$ \ti{cannot}
be turned into an inductive invariant by adding more clauses. In
particular, the broken pushing condition cannot be fixed for $C$.  The
second reason for breaking the pushing condition is that clause $C$
excludes a state \pnt{s} that is \ti{unreachable} in $(k+1)$-th time
frame from an initial state.  In this case, every $F_k$-state
\pnt{q} that is a predecessor of \pnt{s} can be excluded by deriving
a clause falsified by \pnt{q}. So in this case, the broken pushing
condition \ti{can} be fixed. In particular, by fixing broken pushing
conditions for $F_k$ one may turn the latter into an inductive
invariant.

\section{Introducing \icf}
\label{sec:fix_prob}
%
%
\subsection{A high-level view of \icf}
\label{ssec:high_level}
We will refer the version of \ict with a better convergence rate
described in this paper as \tb{\ti{IC4}}.  The main difference between
\ict and \icf is that the latter makes an extra effort in pushing
clauses to later time frames. This new feature of \icf is implemented
in a procedure called \ti{NewPush} (see Figure~\ref{fig:new_push}). It
is invoked after \icf has built $F_k$ where the predecessors of bad
states are excluded i.e. as soon as $F_k \wedge T \rightarrow P'$
holds. For every clause $C$ of $F_k$, \ti{NewPush} checks the pushing
condition (see Section~\ref{sec:conv_prob}). If this condition is
broken, \ti{NewPush} tries to fix it or proves that it cannot be fixed
and hence $C$ is ``unpushable''.

Depending on the clause-pushing effort, one can identify three
different versions of \icf : minimal, maximal and heuristic. The
\ti{minimal} \icf stops fixing pushing conditions as soon as
\ti{NewPush} finds a clause of $F_k$ that cannot be pushed. After that
the minimal \icf switches into the ``\ict mode'' where the pushing
conditions are not fixed for the remaining clauses of $F_k$. The
\ti{maximal} \icf tries to fix the pushing condition for every
inductive clause of $F_k$. That is if a clause $C \in F_k$ cannot be
pushed to $F_{k+1}$, the maximal \icf tries to fix the pushing
condition (regardless of how many unpushable clauses of $F_k$ has been
already identified). Moreover, if an inductive clause $C$ is added to
$F_i$, $i < k$, the maximal \icf try to fix the pushing condition for
$C$ if it cannot be immediately pushed to $F_{i+1}$.

A \ti{heuristic} \icf uses a heuristic to stay between minimal and
maximal \icf in terms of the clause-pushing effort.  In this paper, we
describe the minimal \icf unless otherwise stated.  So, when we just
say \icf we mean the minimal version of it.

\setlength{\intextsep}{2pt}
\begin{figure}
\small
\begin{tabbing}
aaa\=bb\=cc\= dd\= \kill
// \Abs{F}{k} = \s{F_0,\dots,F_k}; \\
// \\
$\mi{NewPush}(I,T,P,\Abs{F}{k})$\{\\
\tb{\scriptsize{1}}\> $\mi{NewClauses} := \mi{true}$;  \\
\tb{\scriptsize{2}}\> $F_{k+1} := P$ \\
\tb{\scriptsize{3}}\> while ($\mi{NewClauses}$) \{\\
\tb{\scriptsize{4}}\Tt  $\mi{NewClauses} := \mi{false}$;\\
\tb{\scriptsize{5}}\Tt  foreach $C \in (F_k \setminus P)$ \{  \\
\tb{\scriptsize{6}}\ttt   if $(C \in (F_{k+1} \setminus P))$ continue; \\
\tb{\scriptsize{7}}\ttt   \pnt{s} := $\mi{SAT}(F_k \wedge T \wedge \overline{C'})$;    \\
\tb{\scriptsize{8}}\ttt  if ($\pnt{s} = \mi{nil}$) \{\\
\tb{\scriptsize{9}}\tttt   $F_{k+1} := F_{k+1} \cup \s{C}$ \\
\tb{\scriptsize{10}}\tttt   continue; \}  \\ 
\tb{\scriptsize{11}}\ttt $(\Abs{F}{k},\pnt{t}) := \mi{ExclState}(\pnt{s},I,T,P,\Abs{F}{k})$;\\
\tb{\scriptsize{12}}\ttt if ($\pnt{t} \neq \mi{nil}$) return($C,\pnt{t}$);\\
\tb{\scriptsize{13}}\ttt $\mi{NewClauses} := \mi{true}$\}\} \\
\tb{\scriptsize{14}}\> return($\mi{nil},\mi{nil}$); \} \\ 
\end{tabbing} 
\vspace{-15pt}
\caption{The $\mi{NewPush}$ procedure}
\vspace{7pt}
\label{fig:new_push}
\end{figure}

%
%
\subsection{Description of \ti{NewPush}}
The pseudo-code of \ti{NewPush} is given in Fig.~\ref{fig:new_push}.
At this point \icf has finished generation of $F_k$. In particular, no
bad state can be reached from an $F_k$-state in one transition.
\ti{NewPush} tries to push every inductive clause of $F_k$ to
$F_{k+1}$. If a clause $C \in F_k$ is unpushable, \ti{NewPush} returns
$C$ and a trace \pnt{t} leading to a state falsified by clause $C$.
Trace \pnt{t} proves the unpushability of $C$ and hence the fact that
$F_k$ cannot be turned into an inductive invariant by adding more
clauses. If every clause of $F_k$ can be pushed to $F_{k+1}$, then
$F_k$ is an inductive invariant and \ti{NewPush} returns
(\ti{nil},~\ti{nil}) instead of clause $C$ and trace \pnt{t}.

\ti{NewPush} consists of two nested loops. A new iteration of the
outer loop (lines 3-13) starts if variable \ti{NewClauses} equals
\ti{true}. The value of this variable is set in the inner loop (lines
5-13) depending on whether new clauses are added to $F_k$.  In every
iteration of the inner loop, \ti{NewPush} checks the pushing condition
(line 7) for an inductive clause of $F_k$ that is not in $F_{k+1}$. If
it holds, then $C$ is pushed to $F_{k+1}$.

If the pushing condition fails, an $F_{k+1}$-state \pnt{s} is
generated that falsifies clause $C$. Then \ti{NewPush} tries to check
if \pnt{s} is reachable exactly as \ict does this when looking for a
counterexample. The only difference is that \pnt{s} is a good
state\footnote{Recall that at this point of the algorithm, no bad
  state can be reached from an $F_k$-state in one transition.}.  As we
mentioned above, if \pnt{s} is reachable by a trace \pnt{t},
\ti{NewPush} terminates returning $C$ and \pnt{t}. Otherwise, it sets
variable \ti{NewClauses} to \ti{true} and starts a new iteration of
the inner loop.

\section{Better Convergence Rate of \icf}
\label{sec:bet_conv_rate}
As we mentioned in Section~\ref{sec:overview}, an upper bound on
\Cr{P} is $\gs{P}$.  Below, we show that using procedure \ti{NewPush}
described in Section~\ref{sec:fix_prob} brings the upper bound on
\Cr{P} for \icf down to \di{\ks}. (Note that if property $P$ holds,
$\di{\ks} \leq \gs{P}$.)

Let $F_k$ be a formula for which \ti{NewPush} is called when $k \geq
\di{\ks}$. At this point $F_k \wedge T \rightarrow P'$ holds.  Let
\pnt{s} be a state breaking the pushing condition for a clause $C$ of
$F_k$.  That is \pnt{s} falsifies $C$ (and hence it is not an
$F_k$-state) but is reachable from an $F_k$-state in one transition.

Recall that $F_k$ is an over-approximation of the set of states that
can be reached in at most $k$-transitions. Since \pnt{s} falsifies
$F_k$, reaching it from an initial state of \ks requires at least
$k+1$ transitions. However, this is impossible since $k +1 > \di{\ks}$
and hence state \pnt{s} is unreachable. This means that every
$F_{k}$-state that is a predecessor of \pnt{s} can be excluded by an
inductive clause added to $F_k$. So eventually, \ti{NewPush} will fix
the pushing condition for $C$.  After fixing all broken pushing
conditions for clauses of $F_k$, \ti{NewPush} will turn $F_k$ into an
inductive invariant.

\section{Number Of Reachable States To Generate}
\label{sec:count_states}

The number of generated reachable states depends on which of the three
versions of \icf is considered (see
Subsection~\ref{ssec:high_level}). Let $k$ denote the maximal number
of time frames unfolded by \icf. In the case of the minimal \icf, the
upper bound on the number of reachable states for proving property $P$
is equal\footnote{For every formula $F_i$, $i=1,\dots,k$, \icf generates one reachable
state \pnt{s} falsifying a clause of $F_i$.  To reach \pnt{s}, one
needs to generate a trace of $i$ states. So the number of reachable
states generated for $F_i$ is equal to $i$.  The total number of
reachable states is equal to $1+2+...+k$.
} to $k*(k+1)/2$. For the maximal
\icf, the upper bound is $k*|\mi{Unpush}(F)|$ where $F = F_1 \cup
\dots \cup F_k$ and $\mi{Unpush}(F)$ is the subset of $F$ consisting
of unpushable clauses.  Indeed, an inductive clause $C \in F_i$ is
proved unpushable only once.  This proof consists of a trace to a
state falsified by $F_i$. The length of this trace is equal to $i$ and
hence bounded by $k$. The upper bound for the maximal \icf above is
loose because one assumes that
\begin{itemize}
\item the  length of every trace proving unpushability equals $k$ 
\item two (or more) clauses cannot be proved unpushable by the same
  reachable state.
\end{itemize}

Re-using reachable states can dramatically reduce the total number of
reachable states one needs to generate. For instance, for the minimal
\icf, this number can drop as low as $k$. For the maximal \icf, the
total number of reachable states can go as low as $m + k$ where $m$ is
the total number of reachable states generated to prove the
unpushability of clauses of $\mi{Unpush}(F)$.

\section{A Few Arguments In Favor Of \icf}
\label{sec:args_in_favor}
In this section, we give some arguments in favor of \icf.  The main
argument is given in Subsection~\ref{ssec:quip} where we relate \icf
with a model checker called \qp.  The latter was
introduced\footnote{As we mentioned in the introduction, at the time
  of writing the first version of our paper we were not aware of \qp.}
in~\cite{quip} in 2015. In Subsections~\ref{ssec:weak_prop}
and~\ref{ssec:testing}, we describe a few potential advantages of \icf
that were not discussed in~\cite{quip} (in terms of \qp).
%
%
\subsection{\icf and \qp}
\label{ssec:quip}
As we mentioned in the introduction, \qp makes an extra effort to push
clauses to future time frames. To show that a clause cannot be pushed,
\qp generates a reachable state. Although the premise of \qp is that
the strategy above may lead to a faster generation of an inductive
invariant, this claim has not been justified theoretically.  The
advantage of \qp over \ict is shown in~\cite{quip} in terms of better
run times and a greater number of solved problems. So, no \ti{direct}
experimental data is provided on whether \qp has a better convergence
rate than \ict. (As mentioned in~\cite{quip} and in the first version
of our paper, having at one's disposal reachable states facilitates
construction of better inductive clauses\footnote{By avoiding the
  exclusion of known reachable states, one increases the chance for an
  inductive clause to be a part of an inductive invariant.}.  So one
cannot totally discard the possibility that the performance of \qp is
mainly influenced by this ``side effect''.)  Nevertheless, great
experimental results of \qp is an encouraging sign.

\subsection{Proving weak properties}
\label{ssec:weak_prop}
In this subsection, we argue that \icf should have more robust
performance than \ict on weak properties.
Let $F_i$ be an over-approximation of the set of states reachable in
at most $i$ transitions and $P$ be the property to prove. As we
mentioned earlier, there are two conditions one needs to satisfy to
turn $F_i$ into an inductive invariant: $F_i \wedge T \rightarrow P'$
and $F_i \wedge T \rightarrow F'_i$. We will refer to a state \pnt{s}
breaking the first condition (respectively second condition) as a
state of the first kind (respectively second kind).  Only states of
the first kind (i.e. $F_i$-states from which there is a transition to
a bad state) are \ti{explicitly} excluded by \ict. States of the
second kind are excluded \ti{implicitly} via generalization of
inductive clauses. On the other hand, \icf excludes states of both
kinds explicitly and implicitly (via generalization of inductive
clauses).

First, assume that $P$ is a \ti{strong} property meaning that there is
a lot of bad states. Then by excluding states of the first kind
coupled with generalization of inductive clauses, \ict also excludes
many states of the second kind.  Now assume that $P$ is a \ti{weak}
property that has, say, only one bad state. Let us also assume that
excluding states reaching this bad state is easy. Intuitively, in this
case, \ict is less effective in excluding the states of the second
kind (because their exclusion is just a \ti{side effect} of excluding
states of the first kind). On the other hand, \icf does not have this
problem and so arguably should have a more robust behavior than \ict
when proving weak properties.

%
%
\subsection{Test generation}
\label{ssec:testing}
Formal verification of \ti{some} properties of transition system \ks
does not guarantee that the latter is
correct\footnote{Moreover, \ks can be incorrect even if a supposedly complete set of
properties $P_1,\dots,P_n$ is proved
true~\cite{cmpl_tst2,fmcad18}. For instance, the designer may
``misdefine'' a property and so instead of verifying the right
property $P'_i$ (that does not hold) a formal tool checks
a \ti{weaker} property $P_i$ (that holds).

}. In this case, testing is employed
to get more confidence in correctness of \ks. Traces generated by \icf
can be used as tests in two scenarios. First, one can check that
reachable states found by \icf satisfy the properties that formal
verification tools failed to prove. Second, one can just inspect the
states visited by \ks and the outputs produced in those states to
check if they satisfy some (formal or informal) criteria of
correctness.

\section{Introducing \Icf}
\label{sec:ic4_pd}
In this section, we present \Icf, a version of \icf employing property
decomposition. In Subsection~\ref{ssec:two_obst}, we describe two
obstacles one has to overcome to make property decomposition
work. Subsection~\ref{ssec:pd_code} introduces a straightforward
implementation of \Icf.
%
%
\subsection{Property decomposition: two obstacles to overcome}
\label{ssec:two_obst}
As we mentioned in the introduction, an appealing feature of \ict is
its ability to generate property-specific proofs. Let $P$ be a hard
property to prove.  Let $P$ be represented as $P_1 \wedge \dots \wedge
P_k$ (i.e. $P$ is decomposed into $k$ weaker properties). Let $J_k$ be
an inductive invariant for property $P_k$. Then $J_1 \wedge \dots
\wedge J_k$ is an inductive invariant for property $P$.  So one can
prove $P$ via finding property-specific proofs $J_i$, $i=1,\dots,k$.

To make the idea of property decomposition work one has to overcome at
least two obstacles. The first obstacle\footnote{This obstacle is of a
  general nature and is not caused by using \ict.} is that the search
space one has to examine to prove $P_i$ is, in general, not a
subset\footnote{The reason is that when proving $P_i$ one may need to
  consider traces that contain two and more
  $\overline{P}$-states. These traces break property $P$ without
  breaking property $P_i$.}  of the search space for
$P$. In~\cite{date2018}, we show that this issue can be addressed by
using the machinery of local proofs\footnote{To prove that $P_i$ holds \ti{globally} one needs to show that no
trace of $P_i$-states reaches a $\overline{P}_i$-state.  Proving $P_i$
\ti{locally} means showing that no trace of $P$-states (rather than
$P_i$-states) reaches a $\overline{P}_i$-state. As we show
in~\cite{date2018}, if $P$ is false, there is property $P_i$ that
breaks both globally and locally. So if every $P_i$ holds locally,
then it does globally too and $P$ is true.
}.

\setlength{\intextsep}{2pt}
\begin{figure}
\small
\begin{tabbing}
aaa\=bb\=cc\= dd\= \kill
$\Icf(I,T,P)$\{\\
\tb{\scriptsize{1}}\> $\mi{Inv} := \emptyset$ \\
\tb{\scriptsize{2}}\> while (\ti{true}) \{ \\
\tb{\scriptsize{3}}\Tt  $\pnt{s} := \mi{CheckSat}(\mi{Inv} \wedge \overline{P})$ \\
\tb{\scriptsize{4}}\Tt  if ($\pnt{s} = \mi{nil}$) return($\mi{Inv},\mi{nil}$) \\
\tb{\scriptsize{5}}\Tt  $Q := FormProp(\pnt{s})$ \\
\tb{\scriptsize{6}}\Tt  $(J,\mi{Cex}) := \mi{IC4}^*(I,T,P,\mi{Inv},Q)$  \\
\tb{\scriptsize{7}}\Tt  if ($\mi{Cex} \neq \mi{nil}$) return($\mi{nil},\mi{Cex}$) \\
\tb{\scriptsize{8}}\Tt if ($J = Q$)   \\
\tb{\scriptsize{9}}\ttt   $J := \mi{Strengthen}(I,T,\mi{Inv},J)$ \\
\tb{\scriptsize{10}}\Tt  $\mi{Inv} := \mi{Inv} \wedge J$ \} \}  \\

\end{tabbing} 
\vspace{-15pt}
\caption{The \Icf procedure}
\vspace{7pt}
\label{fig:ic4_pd}
\end{figure}

The second obstacle is as follows.  As we argued in
Subsection~\ref{ssec:weak_prop}, weak properties are more likely to
expose the convergence rate problem of \ict. For that reason,
replacing a strong property $P$ with weaker properties $P_i$ may
actually lead to performance degradation if properties $P_i$ are
proved by \ict. On the other hand, \icf should be more robust when
solving weak properties. So one can address the second obstacle by
using \icf (rather than \ict) to prove properties $P_i$.
%
%
\subsection{Description of \Icf}
\label{ssec:pd_code}
The pseudocode of \Icf is shown in Fig.~\ref{fig:ic4_pd}.  \Icf
accepts formulas $I,T,P$ specifying the initial states, the transition
relation and the property to prove respectively.  \Icf returns either
an inductive invariant $\mi{Inv}$ or a counterexample $\mi{Cex}$.
Computation is performed in a \ti{while} loop. First, \Icf checks if
there is a $\overline{P}$-state \pnt{s} breaking $\mi{Inv} \rightarrow
P$ (line 3). If not, then \ti{Inv} is an inductive invariant proving
$P$ (line 4). Otherwise, \Icf forms a new property $Q$ to prove (line
5). $Q$ consists of one clause, namely, the longest clause falsified
by \pnt{s}. So, the latter is the only $\overline{Q}$-state.

Then \Icf calls \icfl, a version of \icf that proves $Q$
\ti{locally}\footnote{Proving $Q$ locally addresses the first obstacle mentioned in
Subsection~\ref{ssec:two_obst}. The second obstacle is addressed by
using \icf instead of \ict.

} with respect to the target
property $P$ (see Subsection~\ref{ssec:two_obst}). That is \icfl
checks is there is a trace of $P$-states (rather than $Q$-states)
leading to the $\overline{Q}$-state. If not, then $Q$ holds
locally. \icfl uses the current $\mi{Inv}$ as a constraint\footnote{It
  is safe to do because all reachable states satisfy
  $\mi{Inv}$.}. Namely, \icfl looks for a formula $J$ satisfying
$\mi{Inv} \wedge J \wedge T \rightarrow J'$ (rather than $J \wedge T
\rightarrow J'$).

If \icfl finds a counterexample $\mi{Cex}$, then $Q$ and hence $P$
fail (line 7). Otherwise, \icfl returns an inductive invariant $J$.
If $Q$ is itself an inductive property (and so $J=Q$), \icf tries to
strengthen $J$ like an inductive clause is strengthened by \ict (line
9). This is done to avoid enumerating $\overline{P}$-states one by one
if many properties $Q$ turn out to be inductive. If $J$ is already
strengthened (and so $J \neq Q$), then $\mi{Inv}$ is replaced with
$\mi{Inv} \wedge J$ and a new iteration begins.

\bibliographystyle{plain}
\bibliography{short_sat,local}
\end{document}